\documentclass[a4paper,10pt,onecolumn,oneside,notitlepage,openbib]{article}
\usepackage{amssymb}
\usepackage{amsmath}
\usepackage{latexsym}
\usepackage{array}

\def\DIS{\displaystyle}

\def\qed{\hfill\hbox{$\Box$}\vspace{10pt}\break}
\usepackage{theorem}
\theoremstyle{break}
\newtheorem{Theorem}{Theorem}

\newtheorem{Proposition}{Proposition}
\newtheorem{Lemma}{Lemma}
\newtheorem{Corollary}{Corollary}

\def\ket#1{|#1\rangle}
\def\bra#1{\langle#1|}
\def\C{{\mathbb C}}
\def\R{{\mathbb R}}

\author{
 Fumitaka Yura\thanks{Imai quantum computing and information project,
    ERATO, JST, Daini Hongo White Bldg. 201, 5-28-3 Hongo, Bunkyo,
    Tokyo 113-0033, Japan}
}
\title{Entanglement Cost of Three-Level Antisymmetric States}

\begin{document}
\maketitle

\begin{abstract}
We show that the entanglement cost of the three-dimensional 
antisymmetric states is one ebit.
\end{abstract}

The concept of entanglement is the key for quantum information processing.
To quantify the resource of entanglement, its measures should be
additive, such as bits for classical information.
One candidate for such additive measures is entanglement of formation.
In \cite{Hayden}, it is shown that the entanglement cost $E_c$ to create
some state can be asymptotically calculated from the entanglement of
formation. In this sense, the entanglement cost has an important
 physical meaning.
Since the known results are, nevertheless, not so much
\cite{Vidal,Matsumoto02}, we pay 
attention to antisymmetric states that are easy to deal with.

As is already shown\cite{Shimono03}, the entanglement of formation for
two states in ${\cal S}\left( {\cal H}_{-} \right)$ is additive.
Furthermore, the lower bound for entanglement cost of density matrices
 in $d$-level antisymmetric space, obtained in \cite{Shimono02}, is
  $\log_{2} \frac{d}{d-1}$ ebit.
In this paper, we show that the entanglement cost
of three-level antisymmetric states ($d=3$) in 
${\cal S}\left( {\cal H}_{-} \right)$ is exactly one ebit.

We first define the three-level antisymmetric states.
Let us consider a bipartite qutrit system, ${\cal H}_A = {\cal H}_B = \C^{3}$.
The antisymmetric subspace ${\cal H}_{-}$
on ${\cal H}_A \otimes {\cal H}_B$ is defined as follows:
\[
{\cal H}_{-} := \mbox{span}_{\C} \left\{
\ket{01}-\ket{10}, \ket{12}-\ket{21}, \ket{20}-\ket{02}
\right\} \subset {\cal H}_{A} \otimes {\cal H}_{B}.
\]
Then, the antisymmetric state on ${\cal H}_{-}^{\otimes n}$
 shared with Alice and Bob is, in general,
\begin{eqnarray}
\ket{\psi} & = &
  \sum_{\substack{j_1, j_2, \ldots, j_n = 0\\k_1, k_2, \ldots, k_n=0}}^{2}
\alpha_{j_1, j_2, \ldots, j_n; k_1, k_2, \ldots, k_n}
\ket{j_1, j_2, \ldots, j_n; k_1, k_2, \ldots, k_n} \label{eq:an} \\
& \in &
 {\cal H}_{-}^{\otimes n} \subset
 {\cal H}_A^{(1)} \otimes {\cal H}_A^{(2)} \otimes \cdots
   \otimes{\cal H}_A^{(n)} \otimes
 {\cal H}_B^{(1)} \otimes {\cal H}_B^{(2)} \otimes \cdots
   \otimes{\cal H}_B^{(n)} , \nonumber
\end{eqnarray}
\begin{equation}
\label{eq:defalpha}
\alpha_{j_1, j_2, \ldots, j_n; k_1, k_2, \ldots, k_n} := 
\left( \frac{1}{\sqrt{2}} \right)^{n}
\sum_{i_1, i_2, \ldots, i_n=0}^{2} a_{i_1, i_2, \ldots, i_n}
\prod_{m=1}^{n} \epsilon_{i_m j_m k_m} ,
\end{equation}
where ${\cal H}_{A(B)}^{(i)}$ means $i$th space of
 Alice (resp. Bob) and
$\epsilon$ the Levi-Civita symbol, {\it i.e.}, 
$\epsilon_{ijk}=1$ for $(ijk)=(123)$ and its even permutations,
$-1$ for odd permutations and $0$ otherwise.
Henceforth, we identify the above coefficient
$\alpha_{j_1, \ldots, j_n; k_1, \ldots, k_n}$ with the entries of
 a matrix $\alpha \in M(3^n; \C)$ with respect to the rows
 $\{j_1, \ldots, j_n\}$
 and the columns $\{k_1, \ldots, k_n\}$ with lexicographical order.

The entanglement of formation $E_f$ is defined as follows:
\begin{equation}
\label{eq:ef}
E_f(\rho) = \mbox{inf} \sum_{j} p_j E\left(\ket{\psi_j}\right),
\end{equation}
where $p_j$ and $\ket{\psi_j}$ are decompositions such that 
$\rho = \sum_{j} p_j \ket{\psi_j}\bra{\psi_j}$
 and $E$ is the entropy of entanglement
\begin{equation*}
  E(\ket{\psi}) = S( \mbox{tr}_B \ket{\psi}\bra{\psi} ).
\end{equation*}
The following lemma is well known:
\begin{Lemma}[Subadditivity]
Let $\rho^{(i)}$ be density matrices on 
 ${\cal H}_{A} \otimes {\cal H}_{B}$, {\it i.e.}, bipartite states.
Then,
\begin{equation*}
E_f(\otimes_{i=1}^{n} \rho^{(i)}) \le
  \sum_{i=1}^{n} E_f \left( \rho^{(i)} \right).
\end{equation*}
\end{Lemma}
\begin{Proof}

Let the decomposition for $E_f$ be
\begin{equation*}
  \otimes_{i=1}^{n} \rho^{(i)}
  = \sum_{j} p_j \ket{\psi_j}\bra{\psi_j} \in 
  {\cal S} \left(
    {\cal H}_{A}^{\otimes n} \otimes {\cal H}_{B}^{\otimes n}
  \right)
\end{equation*}
and
\begin{equation*}
  \rho^{(i)} = \sum_{j_i} p_{j_i}^{(i)}
  \ket{\psi_{j_i}^{(i)}}\bra{\psi_{j_i}^{(i)}} \in
  {\cal S} \left( {\cal H}_{A} \otimes {\cal H}_{B} \right)
  \mbox{\ for all $i$}.
\end{equation*}  
\begin{eqnarray*}
\displaystyle
E_f(\otimes_{i=1}^{n} \rho^{(i)}) & = &
  \mbox{inf} \sum_{j} p_j E\left(\ket{\psi_j}\right) \\
& \le &
  \mbox{inf} \sum_{j_1, \ldots, j_n}
  \left( \prod_{i=1}^{n} p_{j_i}^{(i)} \right)
  E\left(\otimes_{i=1}^{n} \ket{\psi_{j_i}^{(i)}} \right) \\
& = &
  \mbox{inf} \sum_{j_1, \ldots, j_n}
  \left( \prod_{i=1}^{n} p_{j_i}^{(i)} \right)
  \sum_{i=1}^{n} E \left( \ket{\psi_{j_i}^{(i)}} \right) \\
& = &
  \mbox{inf} \sum_{i=1}^{n}
  \sum_{j_i} p_{j_i}^{(i)} E \left( \ket{\psi_{j_i}^{(i)}} \right) \\
& = & \sum_{i=1}^{n} E_f \left( \rho^{(i)} \right).
\end{eqnarray*}
\qed
\end{Proof}

Hereafter we use properties of antisymmetric states.
In \cite{Vidal}, it is shown that $E_f(\rho) = 1$
 for any $\rho \in {\cal S} ({\cal H}_{-})$.
Using their result, we obtain the following:
\begin{Corollary}
\label{cor:sub}
For any $\rho^{(i)} \in {\cal S} ({\cal H}_{-})$,
\begin{equation*}
  E_f \left( \otimes_{i=1}^{n} \rho^{(i)} \right) \le n.
\end{equation*}
\end{Corollary}
To prove $E_c = 1$, it is therefore sufficient that we show the 
 superadditivity
$E_f\left(\otimes_{i=1}^{n} \rho^{(i)}\right) \ge n$.
For the states in ${\cal H}_{-}^{\otimes n}$, we can prove the
following lemma:
\begin{Lemma}
\label{lemma:sup}
For any $\ket{\psi} \in {\cal H}_{-}^{\otimes n}$, 
\begin{equation}
\label{eq:evn}
  E \left( \ket{\psi} \right) \ge n.
\end{equation}
\end{Lemma}
We give a proof of this lemma in appendix.
The following corollary immediately follows from this lemma 
because the definition of the entanglement of formation 
 (\ref{eq:ef}) is a linear combination of (\ref{eq:evn}).
\begin{Corollary}
\label{cor:sup}
For any $\rho \in {\cal S} \left( {\cal H}_{-}^{\otimes n} \right)$, 
\begin{equation*}
  E_f \left( \rho \right) \ge n.
\end{equation*}
\end{Corollary}

\begin{Theorem}
For any $\rho^{(i)} \in {\cal S} \left( {\cal H}_{-} \right)$, 
\begin{equation*}
  E_f \left( \otimes_{i=1}^{n} \rho^{(i)} \right) = n.
\end{equation*}
\end{Theorem}
\begin{Proof}
 From the corollaries \ref{cor:sub} and \ref{cor:sup},
 this theorem holds.
\qed
\end{Proof}
Hence, as a corollary of this theorem, we obtain the main result:
\begin{Corollary}[Main Result]
For any $\rho \in {\cal S} \left( {\cal H}_{-} \right)$, 
\begin{equation*}
  E_f \left( \rho^{\otimes n} \right) = n.
\end{equation*}
Therefore, 
\begin{equation*}
  E_c\left( \rho \right)  := 
    \lim_{n \to \infty} \frac{1}{n} E_f\left( \rho^{\otimes n} \right)
     = 1.
\end{equation*}
\end{Corollary}

\subsubsection*{Acknowledgement}
The author is grateful to T. Shimono, K. Matsumoto and H. Fan for
helpful discussions on related their works and seminars.

\subsubsection*{Appendix: Proof of Lemma \ref{lemma:sup}}
It is well known that the entanglement of pure states is defined by
von Neumann entropy of the reduced density matrix
$\rho_A = \mbox{Tr}_B \ket{\psi} \bra{\psi} = \alpha \alpha^{\dagger}$, 
where $\alpha$ is $3^n \times 3^n$ matrix, which is defined in (\ref{eq:an}).
Let $\lambda_i$ be the eigenvalues of $\rho_A$ 
  and its elementary symmetric functions
\begin{eqnarray*}
s_1 & := & \sum_{i} \lambda_i = \mbox{Tr} \rho_{A} = 1 \\
s_2 & := & \sum_{i<j} \lambda_i \lambda_j \\
& \vdots & \\
s_{3^n} & := & \prod_{i} \lambda_i = \mbox{det} \rho_{A} ,
\end{eqnarray*}
the power sum
  $I_{k}(\rho_A) = \sum_{i} \lambda_{i}^{k} = \mbox{Tr} {\rho_A}^{k}$,
respectively.
Notice that $\sqrt{s_2}$ is
 the generalized concurrence\cite{Uhlmann, Rungta, Albererio}.
As we will see later, the value of this generalized concurrence is 
closely related to the entanglement of formation in our case. 

\begin{Proposition}
Let $\alpha$ be the coefficient of $\ket{\psi} \in {\cal H}_{-}^{\otimes n}$
and $\rho_A = \alpha\alpha^{\dagger}$.
Then, 
\begin{equation}
  I_2(\rho_A) \le \frac{1}{2^n}.
\end{equation}
\label{prop:I2}
\end{Proposition}
\begin{Proof}
The calculation of $I_2(\rho_A)$ is lengthy but straightforward.
First, let us choose two rows 
 $J := (j_1, j_2, \ldots, j_n), J' := (j'_1, j'_2, \ldots, j'_n) $
and two columns
 $K := (k_1, k_2, \ldots, k_n), K' := (k'_1, k'_2, \ldots, k'_n) $
for a $2\times2$ minor of matrix $\alpha$.
Since $s_k(\rho_A)$ is equal to the square sum of 
all $k\times k$ minors of $\alpha$ or Gramian, 
we therefore obtain (see, {\it e.g.}, \cite{Fan02})
\begin{eqnarray}
s_2(\rho_A) & = & \frac{1}{4}
\sum^{2}_{
  \substack{
    j_1, j_2, \ldots, j_n = 0 \\
    j'_1, j'_2, \ldots, j'_n = 0 \\
    k_1, k_2, \ldots, k_n = 0 \\
    k'_1, k'_2, \ldots, k'_n = 0
  }
}
\left|
\hspace*{-7mm}
\begin{array}{c}
\alpha_{j_1, \ldots, j_n; k_1, \ldots, k_n}
\alpha_{j'_1, \ldots, j'_n; k'_1, \ldots, k'_n} \\
\hspace{15mm} -
\alpha_{j_1, \ldots, j_n; k'_1, \ldots, k'_n}
\alpha_{j'_1, \ldots, j'_n; k_1, \ldots, k_n}
\end{array}
\right|^2 \nonumber \\
& = & \frac{1}{4} \left( \frac{1}{2^n} \right)^2
\sum_{JJ'KK'} \nonumber \\
& & \left|
  \left( \sum_{p_1, \ldots, p_n = 0}^{2} a_{p_1, \ldots, p_n}
    \prod_{m=1}^{n} \epsilon_{p_m j_m k_m} \right)
  \left( \sum_{p'_1, \ldots, p'_n = 0}^{2} a_{p'_1, \ldots, p'_n}
    \prod_{m'=1}^{n} \epsilon_{p'_{m'} j'_{m'} k'_{m'}} \right)
    \right. \nonumber \\
& & - \left.
  \left( \sum_{p_1, \ldots, p_n = 0}^{2} a_{p_1, \ldots, p_n}
    \prod_{m=1}^{n} \epsilon_{p_m j_m k'_m} \right)
  \left( \sum_{p'_1, \ldots, p'_n = 0}^{2} a_{p'_1, \ldots, p'_n}
    \prod_{m'=1}^{n} \epsilon_{p'_{m'} j'_{m'} k_{m'}} \right)
\right|^2 \nonumber \\
& = & \frac{1}{2^{2n+2}}
\sum_{JJ'KK'} \left| \sum_{PP'}
  a_{p_1, \ldots, p_n} a_{p'_1, \ldots, p'_n} \right. \nonumber \\
& & \times \left. \left(
    \prod_{m=1}^{n} \epsilon_{p_m j_m k_m} 
    \prod_{m'=1}^{n} \epsilon_{p'_{m'} j'_{m'} k'_{m'}} 
   -\prod_{m=1}^{n} \epsilon_{p_m j_m k'_m}
    \prod_{m'=1}^{n} \epsilon_{p'_{m'} j'_{m'} k_{m'}} 
  \right) \right|^2 \nonumber \\
& = & \frac{1}{2^{2n+1}}
\sum_{JJ'KK'} \sum_{PP'QQ'}
  a_{p_1, \ldots, p_n} a_{p'_1, \ldots, p'_n} 
  a_{q_1, \ldots, q_n}^{*} a_{q'_1, \ldots, q'_n}^{*}
  \nonumber \\
& &
\times \left(
\prod_{m_1} \epsilon_{p_{m_1}j_{m_1}k_{m_1}}
\prod_{m_2} \epsilon_{p'_{m_2}j'_{m_2}k'_{m_2}}
\prod_{m_3} \epsilon_{q_{m_3}j_{m_3}k_{m_3}}
\prod_{m_4} \epsilon_{q'_{m_4}j'_{m_4}k'_{m_4}}
\right. \nonumber \\
& & \label{eq:s2} \hspace{5mm} - \left.
\prod_{m_1} \epsilon_{p_{m_1}j_{m_1}k_{m_1}}
\prod_{m_2} \epsilon_{p'_{m_2}j'_{m_2}k'_{m_2}}
\prod_{m_3} \epsilon_{q_{m_3}j_{m_3}k'_{m_3}}
\prod_{m_4} \epsilon_{q'_{m_4}j'_{m_4}k_{m_4}}
\right),
\end{eqnarray}
where we denote
  $\displaystyle 
  \sum_{P} \equiv \sum^{2}_{p_1, p_2, \ldots, p_n = 0}$ and 
  $a_P \equiv a_{p_1, p_2, \ldots, p_n}$, {\it etc.},
  for simplicity.
Let us divide (\ref{eq:s2}) into two parts:
\begin{enumerate}
\item First Term
\begin{eqnarray*}
\sum_{JJ'KK'} \left(
\prod_{m_1=1}^{n} \epsilon_{p_{m_1}j_{m_1}k_{m_1}}
\prod_{m_2=1}^{n} \epsilon_{p'_{m_2}j'_{m_2}k'_{m_2}}
\prod_{m_3=1}^{n} \epsilon_{q_{m_3}j_{m_3}k_{m_3}}
\prod_{m_4=1}^{n} \epsilon_{q'_{m_4}j'_{m_4}k'_{m_4}} \right)
\hspace{-12cm} & & \nonumber \\
& = &
\sum_{j_2, \ldots, j_n=0}^{2} \sum_{J'KK'}
\left( 
  \sum_{j_1=0}^{2} \epsilon_{p_1 j_1 k_1} \epsilon_{q_1 j_1 k_1}
\right) \nonumber \\
& &  
 \times \left(
 \prod_{m_1=2}^{n} \epsilon_{p_{m_1}j_{m_1}k_{m_1}}
 \prod_{m_2=1}^{n} \epsilon_{p'_{m_2}j'_{m_2}k'_{m_2}}
 \prod_{m_3=2}^{n} \epsilon_{q_{m_3}j_{m_3}k_{m_3}}
 \prod_{m_4=1}^{n} \epsilon_{q'_{m_4}j'_{m_4}k'_{m_4}} \right) \\
& = & \sum_{K} \left[ \prod_{m=1}^{n} \left(
  \delta_{k_m k_m}\delta_{p_m q_m} - \delta_{k_m p_m}\delta_{k_m q_m}
  \right) \right] \\
& & \ \times
\sum_{K'} \left[ \prod_{m=1}^{n} \left(
  \delta_{k'_m k'_m}\delta_{p'_m q'_m} - \delta_{k'_m p'_m}\delta_{k'_m q'_m}
  \right) \right] \\
& = & 2^{2n} \prod_{m=1}^{n} \delta_{p_m q_m} \delta_{p'_m q'_m} ,
\end{eqnarray*}
where we use the relation
  $\DIS \sum_{j_1=0}^{2} \epsilon_{p_1 j_1 k_1} \epsilon_{q_1 j_1 k_1}
  = \delta_{k_1 k_1}\delta_{p_1 q_1} - \delta_{k_1 p_1}\delta_{k_1 q_1}$.
\item Second Term
\begin{eqnarray*}
\sum_{JJ'KK'} \left(
\prod_{m_1=1}^{n} \epsilon_{p_{m_1}j_{m_1}k_{m_1}}
\prod_{m_2=1}^{n} \epsilon_{p'_{m_2}j'_{m_2}k'_{m_2}}
\prod_{m_3=1}^{n} \epsilon_{q_{m_3}j_{m_3}k'_{m_3}}
\prod_{m_4=1}^{n} \epsilon_{q'_{m_4}j'_{m_4}k_{m_4}} \right)
\hspace{-12cm} & & \\
& = &
\sum_{j_2, \ldots, j_n=0}^{2} \sum_{J'KK'}
\left( 
  \sum_{j_1=0}^{2} \epsilon_{p_1 j_1 k_1} \epsilon_{q_1 j_1 k'_1}
\right) \nonumber \\
& & \times \left(
\prod_{m_1=2}^{n} \epsilon_{p_{m_1}j_{m_1}k_{m_1}}
\prod_{m_2=1}^{n} \epsilon_{p'_{m_2}j'_{m_2}k'_{m_2}}
\prod_{m_3=2}^{n} \epsilon_{q_{m_3}j_{m_3}k'_{m_3}}
\prod_{m_4=1}^{n} \epsilon_{q'_{m_4}j'_{m_4}k_{m_4}} \right) \\
& = & \sum_{KK'} \prod_{m=1}^{n}
  \left(
    \delta_{k_m k'_m}\delta_{p_m q_m} - \delta_{k_m p_m}\delta_{k'_m q_m}
   \right)
  \left(
    \delta_{k'_m k_m}\delta_{p'_m q'_m} - \delta_{k'_m p'_m}\delta_{k_m q'_m}
   \right)
    \\
& = & \prod_{m=1}^{n} 
  \left(
    \delta_{p_m q_m}\delta_{p'_m q'_m} + \delta_{p'_m q_m}\delta_{p_m q'_m}
   \right) .
\end{eqnarray*}
\end{enumerate}
We summarize these terms and obtain the following.
\begin{eqnarray*}
s_2(\rho_A) & = & 
\frac{1}{2^{2n+1}} \sum_{PP'QQ'} a_P a_{P'} a_{Q}^{*} a_{Q'}^{*} \\
& & \times \left[
2^{2n} \prod_{m=1}^{n} \delta_{p_m q_m} \delta_{p'_m q'_m}
- \prod_{m=1}^{n} 
  \left(
    \delta_{p_m q_m}\delta_{p'_m q'_m} + \delta_{p'_m q_m}\delta_{p_m q'_m}
  \right)
\right] \\
& = &
\frac{1}{2} -
\frac{1}{2^{2n+1}} \sum_{PP'QQ'} a_P a_{P'} a_{Q}^{*} a_{Q'}^{*} 
  \prod_{m=1}^{n} 
  \left(
    \delta_{p_m q_m}\delta_{p'_m q'_m} + \delta_{p'_m q_m}\delta_{p_m q'_m}
  \right),
\end{eqnarray*}
and
\begin{eqnarray*}
I_2(\rho_A) & = & s_1(\rho_A)^2 - 2 s_2(\rho_A) \\
& = &
\frac{1}{2^{2n}} \sum_{PP'QQ'} a_P a_{P'} a_{Q}^{*} a_{Q'}^{*} 
  \prod_{m=1}^{n} 
  \left(
    \delta_{p_m q_m}\delta_{p'_m q'_m} + \delta_{p'_m q_m}\delta_{p_m q'_m}
  \right) \\
& = &
\frac{1}{2^{2n}} \sum_{PP'QQ'}
  \prod_{m=1}^{n} 
  \left(
    \delta_{p_m q_m}\delta_{p'_m q'_m} + \delta_{p'_m q_m}\delta_{p_m q'_m}
  \right) \\
& & \ \times \frac{1}{2}
  \left[
    - \left| a_P a_{P'} - a_Q a_{Q'} \right|^2
    + \left| a_P a_{P'} \right|^2
    + \left| a_Q a_{Q'} \right|^2
  \right] \\
& = &
\frac{1}{2^n} - \frac{1}{2^{2n+1}} \sum_{PP'QQ'}
  \prod_{m=1}^{n} 
  \left(
    \delta_{p_m q_m}\delta_{p'_m q'_m} + \delta_{p'_m q_m}\delta_{p_m q'_m}
  \right)
  \left| a_P a_{P'} - a_Q a_{Q'} \right|^2 \\
& \le & \frac{1}{2^n} .
\end{eqnarray*}
We have thus proved the proposition \ref{prop:I2}.     
\qed
\end{Proof}

The following theorem is important:
\begin{Theorem}[Furuta; Special case of \cite{Furuta93,Furuta97}]
Let $A$ be invertible positive operator. Then for any positive $x \in \R$
\begin{equation*}
  \label{eq:furuta}
  -A \log_2 A \ge (1-\log_2 x) A - \frac{1}{x} A^2.
\end{equation*}
\end{Theorem}
For hermitian matrix $A$, zero eigenvalues do not affect the
above theorem due to $0\log 0 = 0$.
\begin{Corollary}
Let $S(A) = -\mbox{Tr} (A \log_2 A)$ and $\rho_A$ a normalized density matrix
(i.e. $\mbox{Tr} \rho_A = 1$).
Then
\begin{equation*}
\label{eq:lb}
S(\rho_A) \ge -\log_2 I_2(\rho_A).
\end{equation*}
\end{Corollary}
Hence, $S(\rho_A) \ge n$ and 
this ends the proof of Lemma \ref{lemma:sup}.
\qed

\end{document}